\documentclass[onecolumn]{rsauthor} 
\usepackage{graphicx}
\usepackage{natbib}

 \renewenvironment{thebibliography}[1]{%
   \begin{oldthebibliography}{#1}%
     \setlength{\parskip}{0ex}%
     \setlength{\itemsep}{0ex}%
 }%
 {%
   \end{oldthebibliography}%
 }
\bibfont{\small}
\def\hh{H$_2$}
\def\hho{H$_2$O}
\def\nn{N$_2$}
\def\hp{H$^+$}
\def\hhhp{H$_3^+$}
\def\hhdp{H$_2$D$^+$}
\def\ddhp{D$_2$H$^+$}
\def\ohp{OH$^+$}
\def\hhop{H$_2$O$^+$}
\def\hhhop{H$_3$O$^+$}
\def\hnop{H$_n$O$^+$}
\def\nndp{N$_2$D$^+$}
\def\nnhp{N$_2$H$^+$}
\def\dcop{DCO$^+$}
\def\hcop{HCO$^+$}
\def\mic{$\mu$m}
\def\kms{km\,s$^{-1}$}
\def\kkms{K\,km\,s$^{-1}$}
\def\gtsim{{_>\atop{^\sim}}}
\def\ltsim{{_<\atop{^\sim}}}
\def\rcm{cm$^{-1}$}
\def\scm{cm$^{-2}$}
\def\ccm{cm$^{-3}$}
\def\pow#1#2{#1$\times$10$^{#2}$}
\def\new#1{#1}
\begin{document}
\thispagestyle{empty}
\jname{Phil. Trans. R. Soc. A.}
\artnum{XXXXXXXX}
\markboth{F.F.S. van der Tak}{Molecular \new{probes of} star-forming regions} 
\title{Using deuterated \hhhp\ and other molecular \new{species} to understand the formation of stars and planets} 
\author{F.F.S. van der Tak}
\address{SRON Netherlands Institute for Space Research, Landleven 12, 9747 AD Groningen, The Netherlands} 
\abstract{
The \hhhp\ ion plays a key role in the chemistry of dense interstellar gas clouds where stars and planets are forming. 
The low temperatures and high extinctions of such clouds make direct observations of \hhhp\ impossible, but lead to large abundances of \hhdp\ and \ddhp\ which are very useful probes of the early stages of star and planet formation.
Maps of \hhdp\ and \ddhp\ pure rotational line emission toward star-forming regions show that the strong \new{deuteration} of \hhhp\ is the result of near-complete molecular depletion of CNO-bearing molecules onto grain surfaces, which quickly disappears as cores warm up after stars have formed.

In the warmer parts of interstellar gas clouds, \hhhp\ transfers its proton to other neutrals such as CO and \nn, leading to a rich ionic chemistry. The abundances of such species are useful tracers of physical conditions such as the radiation field and the electron fraction. Recent observations of HF line emission toward the Orion Bar imply a high electron fraction, and we suggest that observations of \ohp\ and \hhop\ emission may be used to probe the electron density in the nuclei of external galaxies.
} 
\date{13 February 2012} 
\maketitle 

\section{Introduction}
\label{s:intro}

The \hhhp\ ion plays a central role in the physics and chemistry of interstellar gas clouds where stars and planets are forming \new{\citep{watson1973}}. In cold parts of such clouds, where temperatures \new{are} $\ltsim$20\,K, \hhhp\ is converted into its deuterated forms, which become the dominant forms of \hhhp\ at $T \ltsim 10$\,K.
Here, \hhhp\ can be studied directly by rotational spectroscopy of \hhdp\ and \ddhp\ in the far-infrared and submillimeter. In warm interstellar clouds ($T\sim$30-100\,K), \new{a} direct study of \hhhp\ is possible through vibrational spectroscopy in the mid-infrared, and this technique is widely used for lines of sight through the diffuse interstellar medium (see contributions by McCall, Indriolo and Geballe to this volume). To peer into highly opaque star-forming regions and to map their spatial structure, emission from indirect probes such as \hcop\ and \nnhp\ is used, which form by proton transfer of \hhhp\ to CO and \nn, \new{respectively}.

The goal of this paper is to review recent work on the astrophysical use of \hhhp\ and its deuterated forms.
Since the role of \hhhp\ in the warm and cold regimes is so different, this review discusses them separately: \S~\ref{s:cold} covers the study of \hhdp\ and \ddhp\ in cold clouds, while \S\S~\ref{s:warm} and \ref{s:alt} treat probes of \hhhp\ in warm regions. Space limitations prevent a complete coverage of all work since the previous \hhhp\ conference in 2006; therefore the \new{content} of this review \new{is} necessarily biased toward the interests of its author. 
In particular, this paper does not cover the recent results on the ortho/para ratios of \hh\ and \hhdp, which may constrain the ages of pre-stellar cores; see the contribution by Pagani to this volume. 
This work is also of interest \new{as} it marks the first use of the detailed state-to-state rate coefficients for de-excitation of \hhdp\ in collisions with \hh\ \citep{hugo2009}. Collisions between \hhhp\ and \hh\ may lead to reaction (fully elastic case), excitation (fully inelastic case), or both. 
Thus, the rates by Hugo et \new{al.} are a major step forward for the interpretation of interstellar \hhdp\ lines, compared to the scaled radiative rates used previously (see discussion by \citealt{vandertak2005}). The next step up from Hugo's statistical approach would be full quantum scattering calculations, which would be a large effort because of the reactive nature of deuterated \hhhp.

Another important topic that this paper does not discuss in detail are observations of \hhdp\ in protoplanetary disks. The conditions in the midplanes of such disks are somewhat similar to those in pre-stellar cores: the densities are high enough ($\gtsim$10$^8$\,\ccm) that ultraviolet radiation does not penetrate, leading to a low temperature ($<$10\,K) and a low ionization fraction. The theory of chemistry in the limit of complete depletion of CNO-bearing species \citep{walmsley2004} was originally developed to describe pre-stellar cores, but may be even more applicable to protoplanetary disk midplanes. Here, all but the lightest molecular species freeze out onto grain surfaces, and \hhdp\ is the only available ion to probe the ionization rate of the gas, which controls the efficiency of interaction with the magnetic field. Since the magneto-rotational instability is thought to play a major role in the dynamics of such disks, several teams have searched for \hhdp\ line emission. After an initial \new{claimed detection} by \citet{ceccarelli2004}, \new{only} upper limits have been reported by \citet{guilloteau2006}, \citet{qi2008}, \citet{chapillon2011} and \citet{oeberg2011}. See these papers for the quantitative implications of the limits, and the perspectives with future telescopes.

\section{Cold gas: Spatial distribution of \hhdp\ and \ddhp}
\label{s:cold}

As reviewed previously \citep{vandertak2006}, the astrophysical use of deuterated \hhhp\ is based on the fact that the formation of \hhdp\ and \ddhp\ from \hhhp\ + HD is strongly enhanced at low temperatures ($\ltsim$20\,K). As a result, the abundance of deuterated \hhhp\ peaks in the dense, cold concentrations of gas within clouds known as cloud cores. The temperature dependence is enhanced by a second effect: the main destroyer of deuterated \hhhp\ is CO, which freezes out onto grain surfaces under \new{the same} dense and cold conditions. These two effect make \hhdp\ and \ddhp\ ideal tracers of the initial conditions of star formation; see \citet{sipilae2010} for recent models of this situation. At higher temperatures ($\sim$20--30\,K), reactions with CH$_2$D$^+$ and C$_2$HD$^+$ rather than \hhdp\ act to enrich molecular D/H ratios \citep{parise2009,sakai2009,oeberg2012}.

The importance of deuterated \hhhp\ has been realized for a long time \new{e.g., \citealt{pagani1992}} but whether this potential could be used remained unclear until the first detection spectra of interstellar \hhdp\ \citep{stark1999,caselli2003}.
Motivated by these results, \citet{caselli2008} present the first survey of \hhdp\ line emission, toward a sample of 10 starless cores and 6 protostellar cores. Using the Caltech Submillimeter Observatory,
the \hhdp\ line at 372\,GHz is detected in 7 starless cores and in 4 protostellar cores. The column density of o-\hhdp, derived from the 372\,GHz line intensity, is also higher in starless cores than in protostellar cores. The brightest \hhdp\ lines are detected toward the densest and most centrally concentrated starless cores, where the CO depletion factor (measured by the ratio of CO line emission to dust continuum emission) and the deuterium fractionation (measured as the \nndp/\nnhp\ line ratio) are also largest.

The results of this survey are consistent with the expectation that \hhdp\ traces very dense and cold gas, partly because the reaction of \hhhp\  + HD proceeds mainly in the forward direction, and partly because the main destroyer of \hhdp,CO, freezes out onto grain surfaces.
In particular, at gas temperatures above 15 K, low CO depletion factors and large abundance of negatively charged small dust grains or PAHs drastically reduce the deuterium fractionation to values inconsistent with those observed toward pre-stellar and protostellar cores. 

Comparing their results to a chemical model, \citet{caselli2008} find that several parameters, including the cosmic-ray ionization rate, the depletion of CO \new{(and generally of neutral species)}, the volume density, the dust grain size distribution and the fraction of PAHs all affect the \hhdp\ abundance. \new{In addition, the ortho-to-para ratio of \hh\ is a key parameter \citep{pagani2009}.}
The current data do not allow to disentangle these effects: observations of the p-\hhdp\ ground state line would help to break this degeneracy, as would detailed measurements of the temperatures and densities of the cores as a function of radius.

The results by \citet{caselli2008} are corroborated by mapping of \hhdp\ line emission in the Oph~B2 core \citep{friesen2010}. The column density distributions of both \hhdp\ and \nndp\ in this core show no correlation with the total \hh\ column density. Instead, the deuterium fractionation in the Oph~B2 core, again measured by the \nndp/\nnhp\ ratio, systematically decreases with proximity to the embedded protostars, out to distances $\gtsim$0.04\,pc.

Using the multi-pixel CHAMP+ receiver on the APEX telescope, \citet{parise2011} have successfully detected \ddhp\ emission at 692\,GHz from the H-MM1 prestellar core located in the LDN 1688 cloud. The emission is detected on several pixels of the CHAMP+ array, hence extended on a scale of at least 40$''$, corresponding to $\sim$4800 AU. This is the first secure detection of interstellar \ddhp, after the tentative detection by \citet{vastel2004}. The column densities of \hhdp\ and \ddhp\ toward this core are similar, which implies strong depletion of CO: model calculations indicate that 90--99\% of CO is frozen on dust grains.

Other maps of \hhdp\ emission will be published soon. The region NGC 1333 IRAS4, where \hhdp\ was first detected, is part of the Spectral Legacy Survey (SLS) at the JCMT \citep{plume2007}, which initially covered the 330--360\,GHz range but was later extended to 373\,GHz. This survey provides images of molecular line emission over a 2$\times$2$'$ field at 15$''$ resolution toward four different star-forming environments. In the case of NGC 1333 IRAS4, the \hhdp\ emission is found to have a spatial distribution unlike that of other ions such as \dcop\ or \nnhp\ \new{or the dust continuum}, which peaks near the embedded protostars. The fact that \hhdp\ emission peaks away from heat sources confirms the idea that \hhdp\ is rapidly destroyed as cores warm up (E. Koumpia et al, in preparation).

The first mapping survey of \hhdp\ toward pre-stellar cores is also underway. Using the JCMT, Di Francesco et al (in preparation) have made maps of \hhdp\ line emission toward a sample of 6 pre-stellar cores, selected from the \nndp/\nnhp\ study by \citet{crapsi2005} to have high \nndp/\nnhp\ ratios. For each of these cores, the \hhdp\ emission is found to peak at the core center. The emission is extended over $\approx$30$''$, as seen in the maps and confirmed by the similarity of \hhdp\ antenna temperatures between the CSO and JCMT telescopes.

\section{Warm gas: Reactive ions in star-forming regions} 
\label{s:warm}

Whereas large abundances of interstellar \hhdp\ and \ddhp\ require low temperatures ($\ltsim$10\,K), proton transfer from \hhhp\ to other molecules also takes place in warm gas. In particular, reactions of interstellar \hhhp\ with species such as CO and \nn\ lead to HCO$^+$ and N$_2$H$^+$ which are widely observed in the interstellar medium. Stable ionic species are useful as tracers of the interaction of interstellar gas with magnetic fields (e.g., \citealt{houde2004}; \citealt{schmid2004}), whereas ions \new{which react rapidly with \hh\ } trace other physical conditions such as the gas density and the local radiation field. The second part of this review paper discusses a few recent results in our understanding of the physics and chemistry of reactive ions in interstellar space.

\subsection{The \hnop\ puzzle}
\label{ss:hnop}

The HIFI instrument is a heterodyne spectrometer onboard ESA's Herschel space observatory, which provides access to the spectral ranges between 480--1250 and 1410--1910 GHz at a resolution of $\approx$0.1~MHz. 
One \new{HIFI highlight} is the discovery of interstellar \ohp\ and \hhop\ \citep{gerin2010,ossenkopf2010}, which are intermediate steps in the ionic formation  route of interstellar \hho. This route begins with charge transfer of \hp\ or \hhhp\ to O, proceeds with reactions of \hh\ with O$^+$, \ohp\ and \hhop, and ends with the dissociative recombination of \hhhop\ which produces \hho. Although this route dominates the formation of \hho\ in the gas phase at temperatures $\ltsim$250~K, which is the bulk part of the neutral interstellar medium, the abundances of \ohp\ and \hhop\ were predicted to be low because these species react rapidly with \hh, which is the bulk species of dense interstellar gas clouds. In other words, the rate-limiting steps of ionic \hho\ formation were thought to be the formation of O$^+$ and the dissociative recombination of \hhhop. 

Observations with HIFI have revealed strong absorption by interstellar \ohp\ and \hhop\ on many lines of sight in our Galaxy \citep{neufeld2010:hnop}, and toward nearby starburst galaxies such as M82 \citep{weiss2010}, NGC 253 and NGC 4945. The large column densities of these reactive species imply that the hydrogen in these clouds is neither purely in atomic form, because no \ohp\ and \hhop\ would be produced, \new{nor} in purely molecular form, because all \ohp\ and \hhop\ would react into \hhhop\ and \hho. Previous models of the structure of molecular clouds, where clouds are either diffuse and hydrogen atomic, or dense and hydrogen molecular, are inconsistent with these observations and must be rejected (e.g., \citealt{snow2006}). 
\new{While PDR models are able to explain the observed amounts of \ohp\ and \hhop\ with diffuse gas where most hydrogen is atomic (Gerin, this volume), the distribution of diffuse and dense gas must be reconsidered, as already indicated by mm-wave absorption line studies of diffuse clouds \citep{lucas1996}.}

A major puzzle in the study of interstellar \ohp\ and \hhop\ (which I collectively call \hnop) is the Herschel-SPIRE spectrum of the active galactic nucleus Mrk~231, where lines of \hnop\ appear in emission \citep{vanderwerf2010}. The large dipole moments and the small reduced masses of \hnop\ imply high line frequencies and large radiative decay rates, so that excitation of their rotational levels requires extremely high densities and line emission is not expected to be observable. Since my previous discussion of this topic at the SMILES conference, also organized by Prof. Oka \citep{vandertak2010}, a second report of \hnop\ emission has appeared for the ultraluminous merger Arp~220 \citep{rangwala2011}. In this galaxy, both the \ohp\ and the \hhop\ lines show a P~Cygni-type profile, which consists of blueshifted absorption and redshifted emission, and which is characteristic of a wind or an outflow.


One hint at the solution of this "\hnop\ puzzle" may be the HF molecule. Like \hnop, HF appears in absorption on most lines of sight through the interstellar medium of our Galaxy \citep{sonnentrucker2010}, which is expected from the large dipole moment and small reduced mass of the molecule. The SPIRE spectrum of Mrk~231 however shows HF in emission \citep{vanderwerf2010} and the SPIRE spectrum of Arp~220 shows a P~Cygni profile for the HF line \citep{rangwala2011}. This behaviour is the same as for \hnop, which suggests (but does not imply) a connection between these species. 


\subsection{Detection of HF emission}
\label{ss:hf}

\begin{figure}[tb]
\centering
\includegraphics[width=7cm,angle=-90]{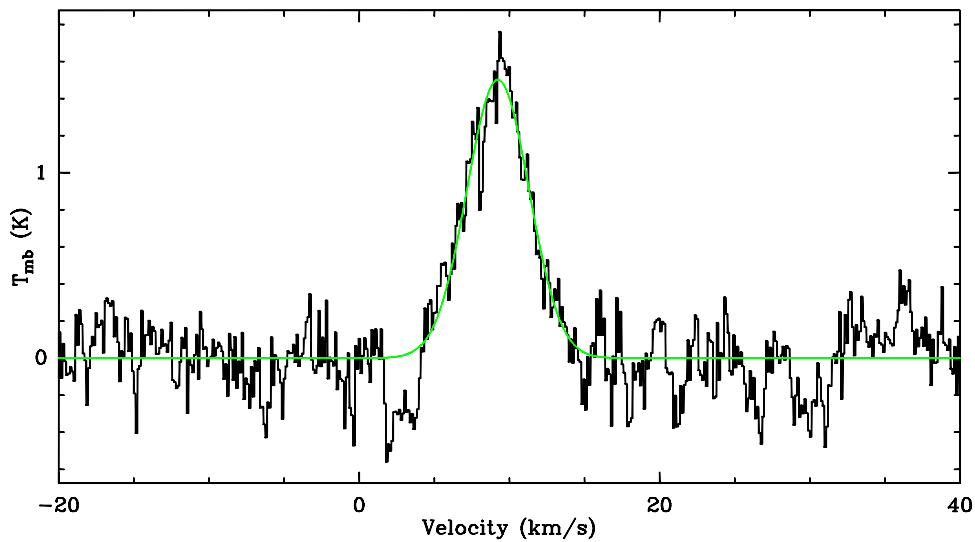}
\caption{Spectrum of the HF $J$=1--0 line, observed with Herschel-HIFI toward the Orion Bar. From: Van der Tak et al (2012).}
\label{f:profi}
\end{figure} 

In dense interstellar gas clouds, the bulk of the hydrogen is in the form of \hh, which is essentially unobservable at low temperatures. The common tracer for \hh\ clouds is the CO molecule, which is abundant and chemically stable, but which photodissociates at \hh\ column densities below $\approx$\pow{1.5}{21}~\scm\ \citep{visser2009}. 
\new{While CH may be used for the intermediate (translucent cloud) regime \citep{sheffer2008}, a proxy for \hh\ at very low column densities was missing until the Herschel mission.}
Since the reaction of F with \hh\ leading to HF is exothermic and HF photodissociation is slow, HF is expected to be the main carrier of fluorine in the gas phase, \new{as long as $\gtsim$10$^{-6}$ of hydrogen is in the form of \hh\ \citep{neufeld2009}}. 

The first confirmation of these predictions came from observations with the LWS instrument onboard the ISO satellite, which measured absorption in the $J$=2--1 line of HF toward Sgr~B2 \citep{neufeld1997}. The Herschel-HIFI instrument gives \new{us} first access to the rotational ground state of HF, and observations of widespread absorption in \new{this} line indicate an HF abundance of \pow{(1--2)}{-8} in diffuse clouds, close to the interstellar (and the solar) abundance of fluorine \citep{sonnentrucker2010}. Absorption is also observed toward dense clouds, but the implied abundance is $\sim$100$\times$ lower \citep{phillips2010}, suggesting that depletion on grain surfaces plays a role.

The large dipole moment of HF and the high frequency of its $J$=1--0 line imply that radiative decay to the ground state is fast, which explains why the line usually appears in absorption on lines of sight through the interstellar medium \citep{neufeld2010:hf,monje2011:sgrb2}. Only very dense environments such as the inner envelopes of late-type stars show the HF line in emission \citep{agundez2011}. The only detection of HF emission from the Galactic interstellar medium so far is toward the Orion Bar, and it is a surprise because the \hh\ density in this region is not high enough to excite the line \citep{vandertak2012}. Instead, collisions with electrons appear to dominate the excitation of HF; non-thermal excitation mechanisms seem less likely, as detailed in \S~\ref{s:alt} of this paper. Line emission of HF thus appears to trace regions of molecular gas with a high electron density, which in the case of the Orion Bar is caused by strong ultraviolet irradiation by the Trapezium stars.


Excitation by electrons may also apply to active galactic nuclei where the Herschel-SPIRE spectra show HF in emission such as Mrk~231 \citep{vanderwerf2010} or as a P~Cygni profile such as Arp~220 \citep{rangwala2011}. In contrast, the Cloverleaf quasar at $z=2.56$ shows HF in pure absorption \citep{monje2011:clover}, which may imply a lower electron density. Given the importance of electron excitation for HF, the appearance of the line in emission or absorption may be a measure of the local ionization rate by ultraviolet photons (in a starburst) or by cosmic rays (in an AGN).

The similarity of the line profiles of HF and \hnop\ in the Herschel-SPIRE spectra of Mrk~231 and Arp~220 may not be a coincidence. Electron impact excitation may also play a role for \hnop\ in these sources. Cross sections for the inelastic collision of the e-\hnop\ system are urgently needed to test this hypothesis. Weak \ohp\ emission is also seen toward the Orion Bar, providing further evidence for this trend (Nagy et al, in prep.).



\section{Alternative mechanisms to produce HF emission}
\label{s:alt}

\subsection{Infrared pumping of HF}
\label{ss:ir-pump}

Radiative excitation of the HF line may occur through the $v$=1--0 vibration mode at 2.55\,\mic. This wavelength is too short for resonances with PAH features, which leaves dust continuum and \hh\ line emission as options. Infrared pumping is not known to play a role for other molecular species in the Orion Bar, but most of these species occur deep in the molecular cloud where near-infrared radiation does not penetrate due to extinction by dust. In contrast, HF readily forms at low extinction where both dust continuum and  \hh\ line emission are strong.

We have calculated the positions of the $v$=1--0 lines of HF from the constants of \citet{ram1996} up to the $J$=10 level. 
Three lines are close to a line of \hh: the P(1) line at 3920.312 \rcm, the R(0) line at 4000.989 \rcm, and the R(1) line at 4038.962 \rcm. Their respective partners are the \hh\ $v$=2--1 Q(1) line at 3920.053 \rcm, the \hh\ $v$=1--0 Q(7) line at 4000.075 \rcm, and the \hh\ $v$=1--0 Q(6) line at 4039.507 \rcm. 
Especially the first match is close, which is encouraging because the P(1) line leads directly to the $J$=1 level observed with HIFI, and the matching \hh\ line is one of the strongest \hh\ lines in Model 14 of \citet{black1987}, with an intensity of $\approx$60\% of the well-known $v$=1--0 S(1) line at 2.12\,\mic. 
However, the separations of these line pairs of 20, 69 and 40 \kms\ exceed the widths of the \hh\ pure rotational lines in the Orion Bar of 4--6\,\kms\ \citep{allers2005}.
The spatial distribution of the \hh\ pure rotational lines is similar to that of the rovibrational lines \citep{vanderwerf1996}, so that their widths are probably similar too; \citet{tielens1993} already rules out a large contribution by shocks to the \hh\ rovibrational line emission from the Orion Bar.

Alternatively, the infrared pumping of HF may take place by continuum radiation from dust. 
Following \citet{carroll1981}, efficient pumping requires that $A_{\rm vib} \epsilon f / (e^{h \nu / kT_d}-1)$ exceeds $A_{\rm rot}$, where $A_{\rm vib}$ and $A_{\rm rot}$ are the vibrational and rotational Einstein A coefficients, $\epsilon$ is the dust emissivity (assumed to be unity in the mid-infrared), $T_d$ is the dust temperature, and $f$ is the geometric filling factor of the dust. Since the vibrational frequency of 3961.4 \rcm\ is high ($h \nu / k$ $>>$ $T_d$), the expression simplifies to $f e^{ -h \nu / kT_d} > A_{\rm rot} / A_{\rm vib}$, where the right hand side is about 1/100 using $A_{\rm vib}$=203\,s$^{-1}$ from \citet{ram1996}.

If the dust radiation field fills the entire sky as seen from the HF cloud ($f$=1), the lower limit for the dust temperature is 1170\,K. While such warm dust could exist locally in the Orion Bar, this temperature seems unreasonably high as an average over a 7200\,AU region. Even transiently heated mini-grains will rarely get so warm. Smaller filling factors exacerbate this problem: for example 50\% sky coverage means a dust temperature of at least 1380\,K, and 10\% coverage implies $T_d$ $>$2340\,K. Hence continuum radiation cannot do the pumping either. 

Finally, we consider the possibility that the excitation of HF in the Orion Bar is influenced by radiation from the Trapezium stars. The brightest of these stars is $\theta^1$C~Ori, which is 127$''$ away from \new{the HIFI} position \citep{vanleeuwen2007}. Assuming that both objects lie 414\,pc from us \citep{menten2007}, the stellar continuum flux at the Orion Bar is 2.6 million times stronger than at the Earth. For the 2.5\,\mic\ flux density, we adopt the K~band magnitude of 4.14 \citep{muench2002} which corresponds to 11.5\,Jy at Earth and 30.3\,MJy at the Bar. Dividing this flux density by 4$\pi$ gives an angle-averaged radiation field $J_\nu$ of \pow{2.4}{-17} erg\,s$^{-1}$\,cm$^{-2}$\,Hz$^{-1}$\,sr$^{-1}$. Setting this equal to the Planck function at a radiation temperature $T_R$, we obtain $T_R = 209$\,K at $\lambda$ = 2.5\,$\mu$m, which is only slightly larger than the intensity due to average Galactic starlight at the Sun's location. The corresponding pumping rate in the HF vibrational fundamental is of order $A_{\rm vib} / \{ \exp(h\nu / kT_R) -1 \}$ = \pow{1.8}{-10}\,s$^{-1}$, which is much less than the collisional excitation rates by \hh\ and electrons. We conclude that infrared pumping does not play a role for HF in the Orion Bar.
 
\subsection{Formation pumping of HF}
\label{ss:chem-pump}

\begin{figure}[p]
\centering
\includegraphics[width=7cm,angle=0]{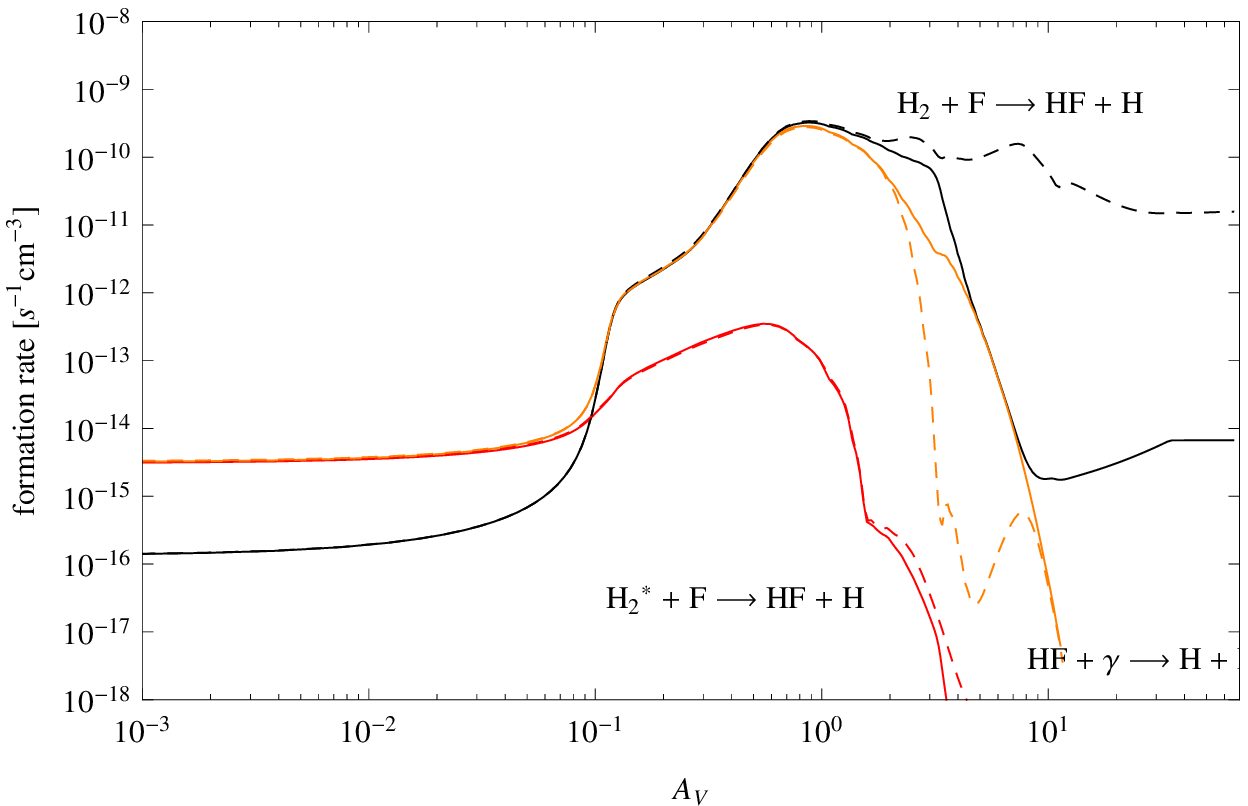}
\includegraphics[width=7cm,angle=0]{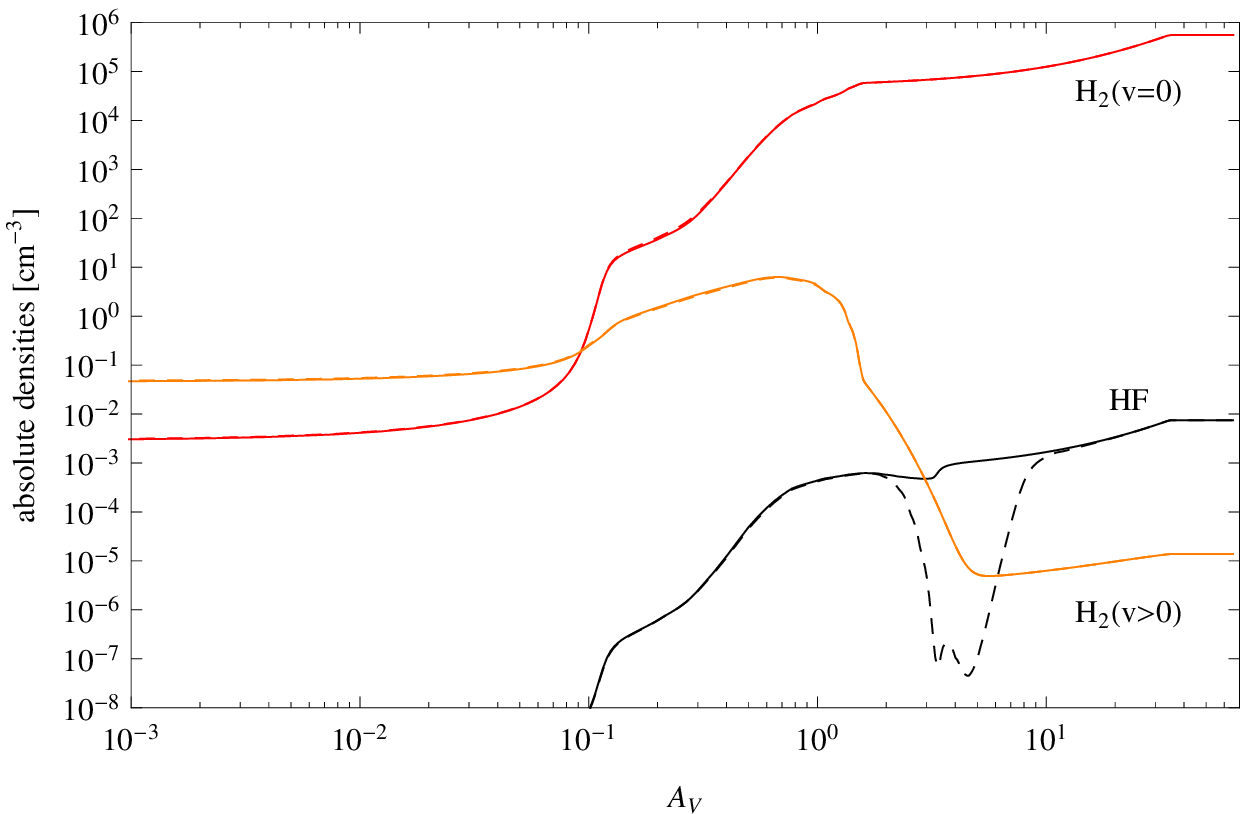}
\includegraphics[width=7cm,angle=0]{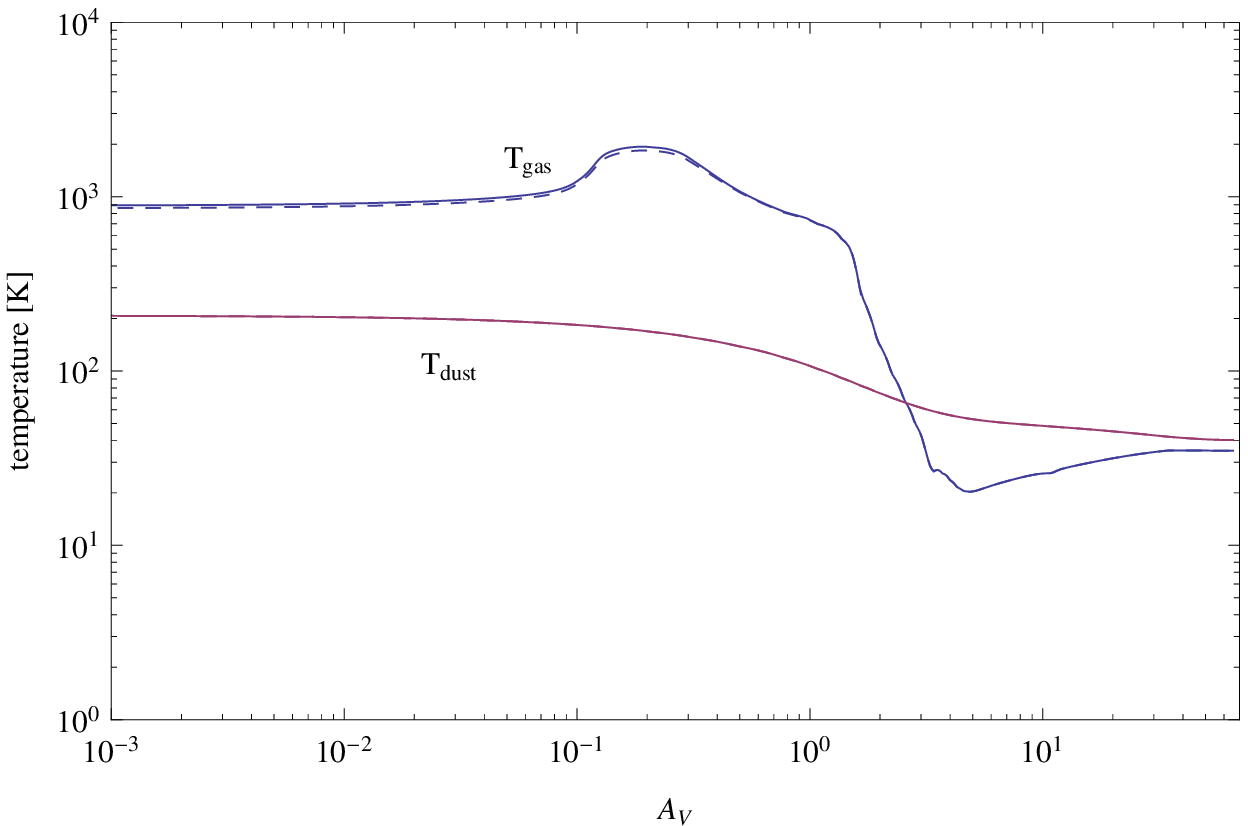}
\caption{Chemical model of the Orion Bar as a function of $A_V$, using the UMIST reaction rates (dashed lines) or the \hh\ + F rate from \citet{neufeld2005} \new{(solid lines)}.
Top: formation rate of HF due to reactions of F with \hh\ in its $v$=0 and $v$=1 states. 
Middle: absolute densities of \hh\ and HF \new{The apparent dip in the HF abundance near $A_V=3$ when using UMIST rates is due to incorrect temperature dependence of the \hh\ + F reaction rate}. 
Bottom: calculated temperatures of gas and dust.
\new{The distribution of HF is unlike that of \ohp\ and \hhop, which for this $\chi$ peak at $0.05<A_V<1$ and disappear at $A_V>1$ where \hh\ and HF are abundant (M. R{\"o}llig, priv. comm.). The observational similarity between \hnop\ and HF is thus likely related to the non-uniform structure of the ISM and similar excitation requirements, rather than chemical similarities.}
}
\label{f:pdr}
\end{figure}

Besides infrared pumping, an alternative non-thermal excitation mechanism of HF would be chemical pumping. 
In this scheme, the excitation of HF is out of thermal equilibrium because the formation and destruction rates are as fast as radiative decay.
This mechanism is plausible for the Orion Bar because of the high ultraviolet radiation field, which leads to a large amount of material in the regime where \hh\ is abundant but the HF destruction rate by photodissociation and reaction with C$^+$ is of the order of the maximum formation rate. 

Assuming optically thin emission, the observed HF line flux of 7.8\,\kkms\ = \pow{1.5}{-5}\,erg\,\scm\,s$^{-1}$\,sr$^{-1}$ requires a column density of $N_{J=1}$ = \pow{9.5}{11}\,\scm\ in the $J$=1 state of HF. To obtain this by chemical pumping, the HF formation rate \new{$R=k n(\mathrm{H_2}) n(\mathrm{F}) / n(\mathrm{HF})$} must fulfil $R/A_{10} \times X_{\rm HF} \times N_H \ge 9.5 \times 10^{11}$\,\scm, where $A_{10}$ is the Einstein A coefficient of the HF $J$=1--0 line. 
The formation of HF is dominated by the reaction of \hh\ with F, the rate coefficient of which is given in Table~2 of \citet{neufeld2009} and equals $k$ = \pow{4.6}{-12}\,cm$^3$\,s$^{-1}$ at $T$ = 80\,K. 
In strong radiation fields, destruction of HF is dominated by photodissociation, at a rate of $\zeta_d$ = \pow{1.2}{-10}\,s$^{-1}$ in the standard Draine field, so that $\zeta_d$ = \pow{6}{-6}\,s$^{-1}$ for the Orion Bar ($\chi \sim$50,000). 
The HF abundance follows from balancing its formation and destruction rates \new{($\zeta_d = R$)}:
\begin{equation}
n({\rm HF}) \zeta_d = k n({\rm H_2}) n({\rm F}) 
\end{equation}
where the sum of the abundances $X({\rm HF})+X({\rm F})$ is constrained by the total amount of fluorine, $1.8\times 10^{-8}$ relative to H.
The product $R/A_{10} \times X($HF$)$ reaches its maximum of \pow{4.5}{-12} for very high densities, leading to a maximum $N_{J=1} = 4.5 \times 10^{11}$~cm$^{-2}$ when assuming
a total gas column density of $10^{23}$~cm$^{-2}$. For gas densities of $10^6$~cm$^{-3}$ this drops by a factor of two, for a gas density of $10^5$~cm$^{-3}$ by another factor of ten. Higher temperatures lower
the required densities approximately proportionally. 

The computation above assumes that every molecule formation leads to an excitation of the $J$=1 state. This premise is reasonable, if not intuitive. The formation of HF releases 1.36~eV in each reaction, which is well above the energy of the $v$=1 state, so that all rotational levels in the vibrational ground state will be about equally populated. The relative fraction of molecules formed in $J$=0 will be very small, and the radiative decay of all other molecules passes through the $J=1$ state so that they contribute to the observed $J$=1--0 emission.

We have calculated the formation rate of HF in a self-consistent model of fluorine chemistry using the KOSMA-\new{$\tau$} code \citep{roellig2006}. 
The rate equations are solved for a spherical clump of gas and dust with a density of $n_H$=10$^5$\,\ccm\ at the outer radius ($A_V$=0) and increasing to \pow{1.1}{6}\,\ccm\ at the clump center, so that the total clump mass is 1\,$M_\odot$. The dust model is number 7 of \citet{weingartner2001}, equivalent to $R_V$=3.1, and the formation of \hh\ is according to \citet{cazaux2002} with formation on PAH surfaces suppressed. 
The total abundance of fluorine in the Orion region is taken from \pow{6.68}{-9} \citep{simon-diaz2011}, which is lower than the average value from \citet{neufeld2009} by about a factor of 3. 

Figure~\ref{f:pdr} shows the results of our calculations, which consider both ground-state and vibrationally excited \hh. Collision of F with \hh\ ({\it v}$>$0) dominates the formation of HF until $A_V$=0.0036, after which point ground-state \hh\ takes over. 
The full model confirms the numbers from the estimate above
with a formation rate $R$ of about $10^{-5}$~s$^{-1}$, 
and approximately half of the fluorine in the form of HF in the
reactive zone between $A_V \approx 0.01 \dots 2$ containing most
of the mass. Hence, we find that chemical pumping of the HF $1-0$
transition is effective in the Orion Bar, but that estimates
based on current molecular data fall short by a factor of a few.
This could be due to uncertainties in the total fluorine abundance,
or higher temperatures and radiation fields in the interclump gas.

\section{Conclusions and future work}
\label{s:future}


The study of interstellar deuterated \hhhp\ has come a long way in the past 16 years, and the conferences that Prof.\ Oka has organized every six years provide clear milestones of this progress. At the 88$^{\rm th}$ birthday of the discovery of \hhhp\ in 2000, the first detection of interstellar \hhdp\ was announced \citep{stark1999}, which followed shortly after the discovery of interstellar \hhhp\ itself \citep{geballe1996}. The 94$^{\rm th}$ birthday saw the first report of strong \hhdp\ emission from cold pre-stellar cores \citep{caselli2003}, 
the first \hhdp\ map of LDN~1544 by \citet{vastel2006}
and the first tentative detection of interstellar \ddhp\ \citep{vastel2004}. This year, at the 100$^{\rm th}$ birthday, we are seeing the confirmation of interstellar \ddhp\ \citep{parise2011}, the first survey of \hhdp\ emission \citep{caselli2008}, and mapping surveys of \hhdp\ are ongoing.

The natural question to ask \new{now is}: What will the 106$^{\rm th}$ birthday in 6 years' time bring? The most likely development in the next years are high-resolution maps of pre-stellar \hhdp\ with the ALMA interferometer, which is now about half-way \new{completed} and has started its Early Science program. In fact, the tuning range of the ALMA receivers was especially extended to include the 372~GHz line of o-\hhdp, following the CSO detection of strong emission in this line in the pre-stellar core LDN~1544. The ALMA telescope will also likely make the first detection of \hhdp\ in a protoplanetary disk, and hence constrain the ionization fraction of the disk midplane. Another possibility is a survey of the p-\ddhp\ line at 692~GHz in pre-stellar cores with ALMA, although the emission is probably quite weak in many cases. Third is the new capability of the GREAT receiver on the SOFIA airborne observatory, which gives access to the p-\hhdp\ ground state line near 1370~GHz. The science operations of SOFIA have started in 2011, and GREAT is among the first-light instruments. 

The GREAT instrument may also be used for searches for the ground-state line of o-\ddhp\ at 1477~GHz. A search for this line with HIFI in the framework of the CHESS program \citep{ceccarelli2010} was unsuccessful (C. Vastel, priv. comm.). Both the p-\hhdp\ 1370~GHz line and the o-\ddhp\ 1477~GHz line are unlikely to appear in emission: their upper level energies of $\sim$70\,K and radiative decay rates of $\sim$\pow{3}{-3}\,s$^{-1}$ imply that their excitation requires higher densities and temperatures than occur in regions where the species are abundant. However, it is possible that the GREAT observations will reveal absorption by p-\hhdp\  and o-\ddhp\ near 1400~GHz. Appearance in absorption requires a source of background continuum, which the warm central regions of young protostellar cores may provide, as well as sufficient abundance of \hhdp\ amd \ddhp\ in the cold outer layers of these cores. Such conditions may well occur along some lines of sight, as the example of \hhdp\ absorption at 372~GHz toward IRAS 16293 shows \citep{stark2004}. These observations would be a new astronomical use of deuterated \hhhp\  to probe physical conditions in star-forming regions.


In warm regions of the interstellar medium, hydride molecules such as HF and reactive ions such as \ohp\ and \hhop\ will continue to play an important role in characterizing the physical conditions of the gas. Mapping the spatial distribution of HF in the Orion Bar and other photon-dominated regions with Herschel-HIFI will help to clarify the contributions by atomic and molecular layers to the HF emission. The Herschel-SPIRE spectra of other galactic nuclei will help us to understand the origin of \ohp\ and \hhop\ emission by probing a range of environments. 
After the end of the Herschel mission in $\approx$March 2013, ground-based observations of \ohp\ such as pioneered by \citet{wyrowski2010} will be invaluable as probes of interstellar physics and chemistry, supported by other reactive ions that are observable from the ground such as SH$^+$ \citep{menten2011}, CO$^+$ \citep{staeuber2009} and HOC$^+$ \citep{rizzo2003}. Clearly, the use of deuterated \hhhp\ and other molecular ions to probe star-forming regions has a bright future!

\ack{I thank M. R\"ollig for input, and \new{L. Pagani and D. Lis} for \new{comments on the manuscript}.}

\begin{small}
\bibliographystyle{aa}
\bibliography{h3+}
\end{small}

\end{document}